\documentstyle[preprint,aps,prb]{revtex}
\begin{document}

\draft

\title{
Quantum Monte Carlo study of the pairing correlation in the Hubbard
ladder
}

\author{
Kazuhiko Kuroki, Takashi Kimura, and Hideo Aoki
}
\address{Department of Physics, University of Tokyo, Hongo,
Tokyo 113, Japan}

\date{\today}

\maketitle

\begin{abstract}
An extensive Quantum Monte Carlo calculation is performed for the 
two-leg Hubbard ladder model to 
clarify whether the singlet pairing correlation decays slowly, 
which is predicted from the weak-coupling theory but 
controversial from numerical studies. 
Our result suggests 
that the discreteness of energy levels in finite systems 
affects the correlation enormously, where 
the enhanced pairing correlation is indeed detected 
if we make the energy levels of the
bonding and anti-bonding bands lie close to each other at the Fermi level 
to mimic the thermodynamic limit.
\end{abstract}

\medskip

\pacs{PACS numbers: 74.20.Mn, 71.10.Fd}

%\begin{multicols}{2}
\narrowtext

\newpage

Over the past several years, strongly correlated 
electrons on ladders 
have received much attention both theoretically and 
experimentally \cite{DagRice}.  
This has been kicked off by the theoretical studies suggesting 
a formation of the spin gap and possible occurrence of
superconductivity in such systems \cite{Dagotto1,Rice}.

The weak-coupling theory with the bosonization and renormalization-group 
techniques \cite{Finkelstein,Fabrizio,Khveshchenko,Schulz,Balents,Nagaosa} 
has indeed shown that the Hubbard model on a two-leg 
ladder has a spin gap, and if the system is free from umklapp processes, 
singlet pairing correlation function decays as
$\sim 1/r^{\alpha}$ with $\alpha=1/2 ( r$: real space distance) 
in the weak-coupling limit.

Since SDW and $2k_F$ CDW correlations have to decay 
exponentially in the presence of a spin gap in a two-leg ladder, 
the only phase competing with 
superconductivity will be $4k_F$ CDW, whose correlation should 
decay as $1/r^{1/\alpha}$. Hence the
pairing correlation dominates over all the others if $\alpha<1$.

As for the opening of the spin gap, 
the density-matrix renormalization group (DMRG) studies 
in the strong-coupling regime also
indicate its presence in both $t$-$J$ and 
Hubbard ladder models\cite{Noack,Noack2,Hayward}. 
If we further focus on the $t$-$J$ ladder, 
DMRG\cite{Hayward} detects a pairing correlation 
decaying slightly slower than $\sim 1/r$ and a CDW correlation decaying
faster than $\sim 1/r$ for an electron density of 
$n=0.8$ with $J/t=1$\cite{diag}. 

However, the dominance of the pairing 
correlation in the {\it Hubbard} ladder model 
seems to be a subtle problem in numerical calculations. 
Namely, a DMRG study by Noack {\it et al.} for the doped Hubbard ladder 
with  $n=0.875$, $U/t=8$, and $t_{\perp}=t$ (where $t$ and $t_{\perp}$ 
are intra- and 
interchain hoppings, respectively) 
shows no enhancement of the pairing correlation
over the $U=0$ result\cite{Noack}, while
they do find an enhancement at $t_{\perp}=1.5t$\cite{Noack2}. 
Asai performed a Quantum Monte Carlo (QMC) calculation for a
36-rung ladder with $n=0.833$, $U/t=2$ and $t_{\perp}=1.5t$\cite{Asai}, 
in which no enhancement of the pairing correlation was found.
On the other hand, Yamaji {\it et al.} have found an enhancement 
for the values of the parameters 
where the lowest anti-bonding band levels for $U=0$ approaches
the highest occupied bonding band levels, although their results 
have not been conclusive due to small system sizes 
($\leq 6$ rungs)\cite{Yamaji}.

Thus, existing analytical and numerical results appear to be 
controversial.  
This is disturbing since 
the superconductivity in the Hubbard ladder, especially with $t_{\perp}\sim t$,
is of great interest as a model for 
cuprate ladder-like materials, for which an occurrence of superconductivity
has indeed been reported very recently\cite{Uehara}. 
In the present work, we have performed an extensive QMC
calculation for the Hubbard ladder with $t_{\perp}\sim t$ in order to 
clarify the origin of the discrepancies among existing results.  
We conclude that 
the discreteness of energy levels in finite systems affects 
the pairing correlation enormously, where 
the enhanced pairing correlation is indeed detected 
if we tune the parameters so as to align the discrete energy levels of
bonding and anti-bonding bands at the Fermi level 
in order to mimic the thermodynamic limit. 

The Hamiltonian of the two-leg Hubbard ladder 
is given in standard notations as 
\begin{eqnarray}
{\cal H}&=&-t\sum_{\alpha i \sigma}
(c_{\alpha i\sigma}^\dagger c_{\alpha i+1\sigma}+{\rm h.c.})\nonumber\\&&
-t_{\perp}\sum_{i \sigma}
(c_{1,i\sigma}^\dagger c_{2,i\sigma}+{\rm h.c.})
+U\sum_{\alpha i} n_{\alpha i\uparrow}n_{\alpha i\downarrow},
\end{eqnarray}
where $\alpha(=1,2)$ specifies the chains.

In the weak-coupling theory, the amplitude of the pair hopping process 
between the bonding and anti-bonding bands in momentum space
flows into the strong-coupling regime upon renormalization,
resulting in a formation of gaps 
in both of the two spin modes and a gap in one of the charge modes
when the umklapp processes are irrelevant.
This leaves one charge mode massless, where the mode is characterized by 
a critical exponent $K_\rho$, which should be close to unity in the 
weak-coupling regime.  Then the correlation function of 
an interchain singlet pairing, 
$O_i=(c_{1i\uparrow} c_{2i\downarrow}-c_{1i\downarrow} c_{2i\uparrow})
/\sqrt{2}$,decays like $1/r^{1/(2K_\rho)}$.  

Here, we have applied the projector Monte Carlo method\cite{pmc}
to look into the ground state correlation function 
$P(r)\equiv \langle O_{i+r}^\dagger O_{i}\rangle$
of this pairing.  
We assume periodic
boundary conditions along 
the chain direction, $c_{N+1} \equiv c_{1}$, 
where $N$ is the number of rungs. 

The details of the QMC calculation are the following.
We took the non-interacting Fermi sea as the trial state.
The projection imaginary time $\tau$ was taken to be $\sim60/t$.
We need such a large $\tau$ to ensure the convergence of especially 
the long-range part of the pairing correlation.  
This sharply contrasts with the situation for single chains, 
where $\tau\sim 20/t$ suffices 
for the same sample length considered here.  
The large value of $\tau$, along with a large 
on-site repulsion $U$, makes the negative-sign problem serious, 
so that the calculation is feasible for $U/t\leq 2$.
In the Trotter decomposition, the imaginary time increment
$[\tau/$(number of Trotter slices)] is taken to be
$\leq 0.1$.  
We have concentrated on
band fillings for which the closed-shell condition 
(no degeneracy in the non-interacting Fermi sea) is met.  
We set $t=1$ hereafter. 

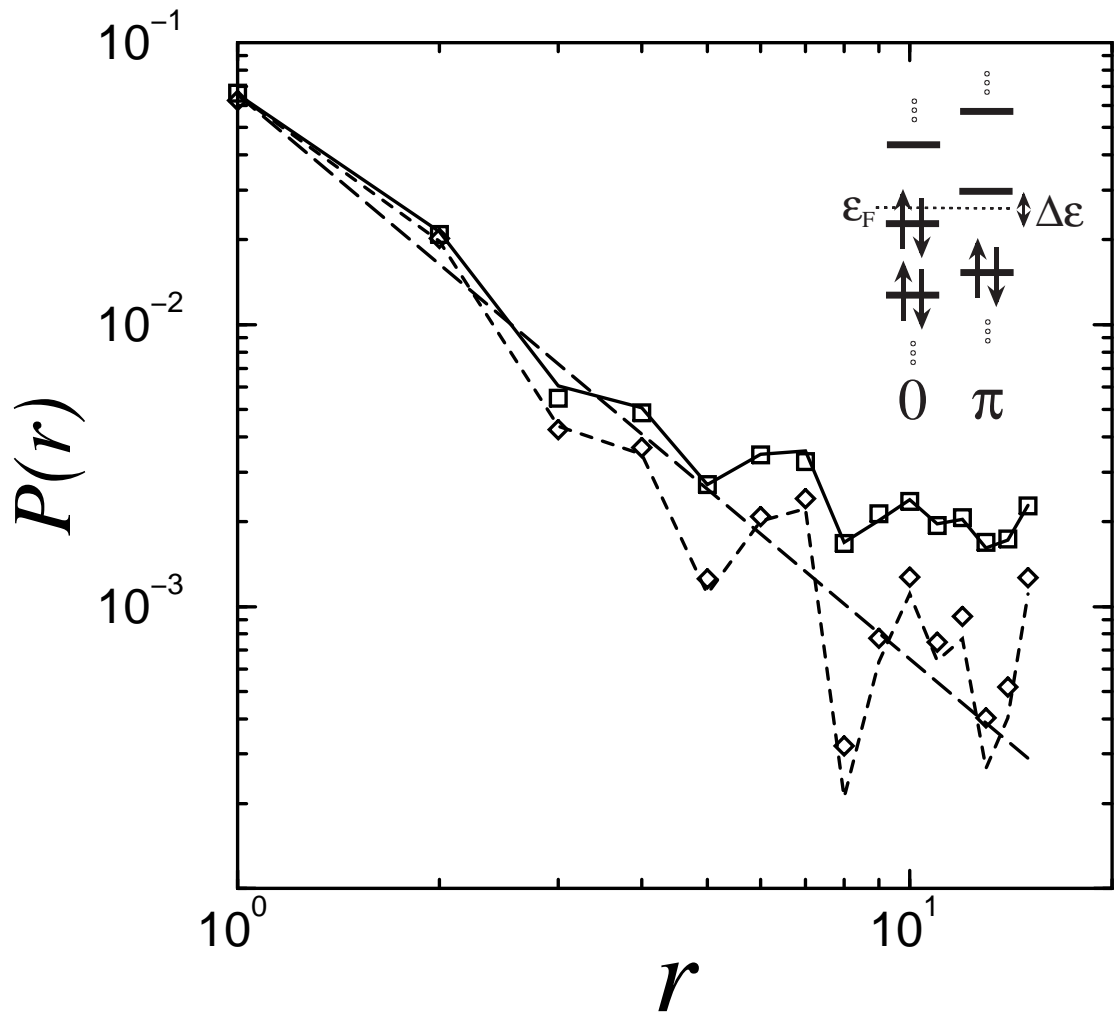
\begin{figure}
\caption{
The pairing correlation function, $P(r)$, plotted 
against the real space distance $r$ in a 30-rung 
Hubbard ladder having 52 electrons 
for $U=1$ with $t_{\perp}=0.98$ 
($\Box$) 
and $t_{\perp}=1.03$ ($\Diamond$). 
The dashed line is the non-interacting
result for the same system size, while the straight dashed line 
represents $\propto 1/r^2$. The solid line is a fit to the 
$U=1$ result with $t_{\perp}=0.98$ (see text). The inset shows a schematic
image of the discrete energy levels of bonding $(0)$ and anti-bonding $(\pi)$
bands for $U=0$.
}
\label{fig1}
\end{figure}

We first show 
in Fig.\ref{fig1} the result for $P(r)$
for $t_{\perp}=0.98$ and $t_{\perp}=1.03$ 
with $U=1$ and 
the band filling $n=0.867=52$ electrons/ (30 rungs $\times$ 2 sites).  
The $U=0$ result (dashed line) 
for these two values of $t_{\perp}$ are identical 
because the Fermi sea remains unchanged.
However, if we turn on $U$, 
the 5\% change in $t_{\perp}=0.98\rightarrow 1.03$ is 
enough to cause a dramatic 
change in the pairing correlation: 
the $t_{\perp}=0.98$ result has a large
enhancement over the $U=0$ result at large distances, while 
the enhancement is not seen for $t_{\perp}=1.03$. 

In fact we have deliberately chosen these values 
to control the alignment of the 
discrete energy levels at $U=0$ in finite, two-band systems.  
Namely,
when $t_{\perp}=0.98$, the one-electron energy levels of the 
bonding and  anti-bonding bands for $U=0$ lie close to each other 
around the Fermi level with the 
level offset ($\Delta\varepsilon$ in the inset of Fig.\ref{fig1}) 
being as small as 0.004, while
they are staggered for $t_{\perp}=1.03$ with the level offset of 0.1.
On the other hand, the size of the spin gap is known to be  
around $0.05t$ for $U=8$\cite{Noack2}, and is expected to be of the  
same order of magnitude or smaller for smaller values of $U$.
The present result then suggests that if the level offset $\Delta\varepsilon$
is too large compared to the spin gap, 
the enhancement of the pairing correlation cannot be seen.
By contrast, for a small enough $\Delta\varepsilon$, by which 
an infinite system is mimicked, the enhancement 
is indeed detected as expected from the weak coupling 
theory, in which the spin gap is assumed to be infinitely large at the fixed 
point of the renormalization flow.

Our result is reminiscent of those obtained by 
Yamaji {\it et al.}\cite{Yamaji}, 
who found an enhancement of the pairing correlation 
in a restricted parameter regime where the lowest anti-bonding 
levels approaches the highest occupied bonding levels.
They conclude that superconductivity occurs 
when the anti-bonding band `slightly touches' the Fermi level. 
However, our result in Fig.\ref{fig1} is obtained for the 
band filling for which no less than seven out of 30 anti-bonding levels 
are occupied at $U=0$.  
Hence the enhancement of the pairing correlation
is not restricted to the situation 
where the anti-bonding band edge touches the Fermi level. 

Now, let us more closely look into the form of $P(r)$ for
$t_{\perp}=0.98$.  It is difficult to determine the decay exponent
of $P(r)$, but here we attempt 
to fit the 
data by assuming a trial function expected from the weak-coupling theory. 
Namely, we have fitted the data with the form
$
P(r)=\frac{1}{4\pi^2}\sum_{d=\pm}\{ cr_d^{-1/2}
+(2-c)r_d^{-2}
-\left[\cos(2k_F^0 r_d)+\cos(2k_F^{\pi} r_d)\right] r_d^{-2}\}
$
with the least-square fit (by taking logarithm of the data) $c=0.11$.  
Because of the periodic boundary condition, 
we have to consider contributions from both ways around, so there are
two distances between the $0$-th and the $r$-th rung, i.e, $r_{+}=r$ and 
$r_{-}=N-r$.
The period of the cosine terms is assumed to be the 
non-interacting Fermi wave numbers of the bonding and the anti-bonding 
bands in analogy with the single-chain case. 
The overall decay 
should be $1/r^2$ as in the pure 1D case.   
We have assumed the form $c/r^{1/2}$
as the dominant part of the correlation at large distances 
because this is what is expected in the weak-coupling theory. 
A finite $U\sim 1$ may give some correction, but the result 
(solid line in Fig.\ref{fig1}) fits to the numerical result 
surprisingly accurately. 
If we least-square fit the 
exponent itself as 
$1/r^\alpha$, we have 
$0.2<\alpha<0.7$ with a similar accuracy.  Thus 
a finite $U$ may change $\alpha$, 
but $\alpha>1$ may be excluded.  
To fit the short-range part of the data, we require 
non-oscillating $(2-c)/r^2$ term, 
which is not present in the weak-coupling theory.  
We believe that 
this is because the weak-coupling theory only concerns with
the asymptotic form of the correlation functions.

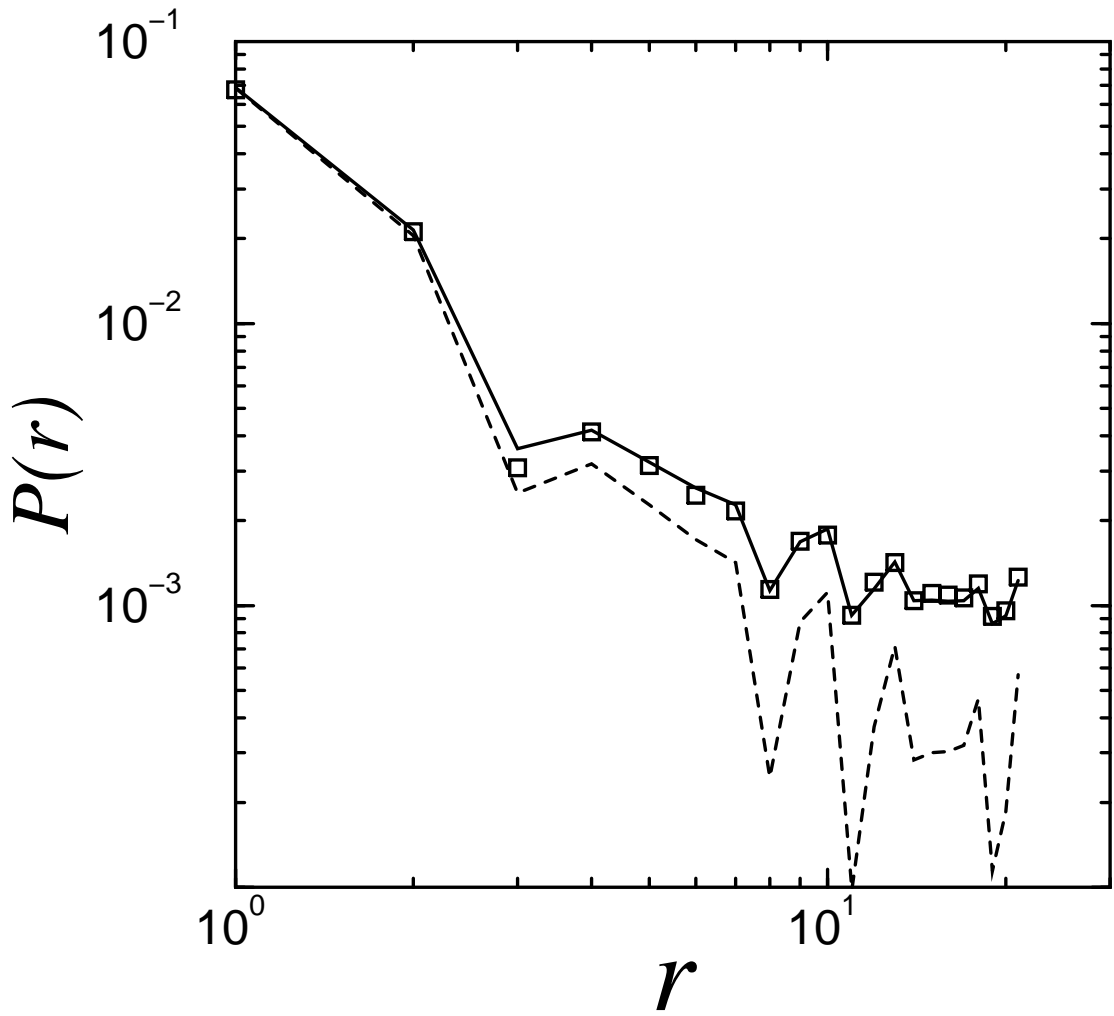
\begin{figure}
\caption{
A similar plot as in Fig.1 for a 42-rung system 
having 76 electrons with $t_{\perp}=0.99$.  
}
\label{fig2}
\end{figure}

In Fig.\ref{fig2}, we show a result for a larger system size (42 rungs) 
for a slightly different electron density, $n=0.905$ with 76 electrons 
and $t_{\perp}=0.99$. 
We have again an excellent fit 
with $c=0.07$ this time.

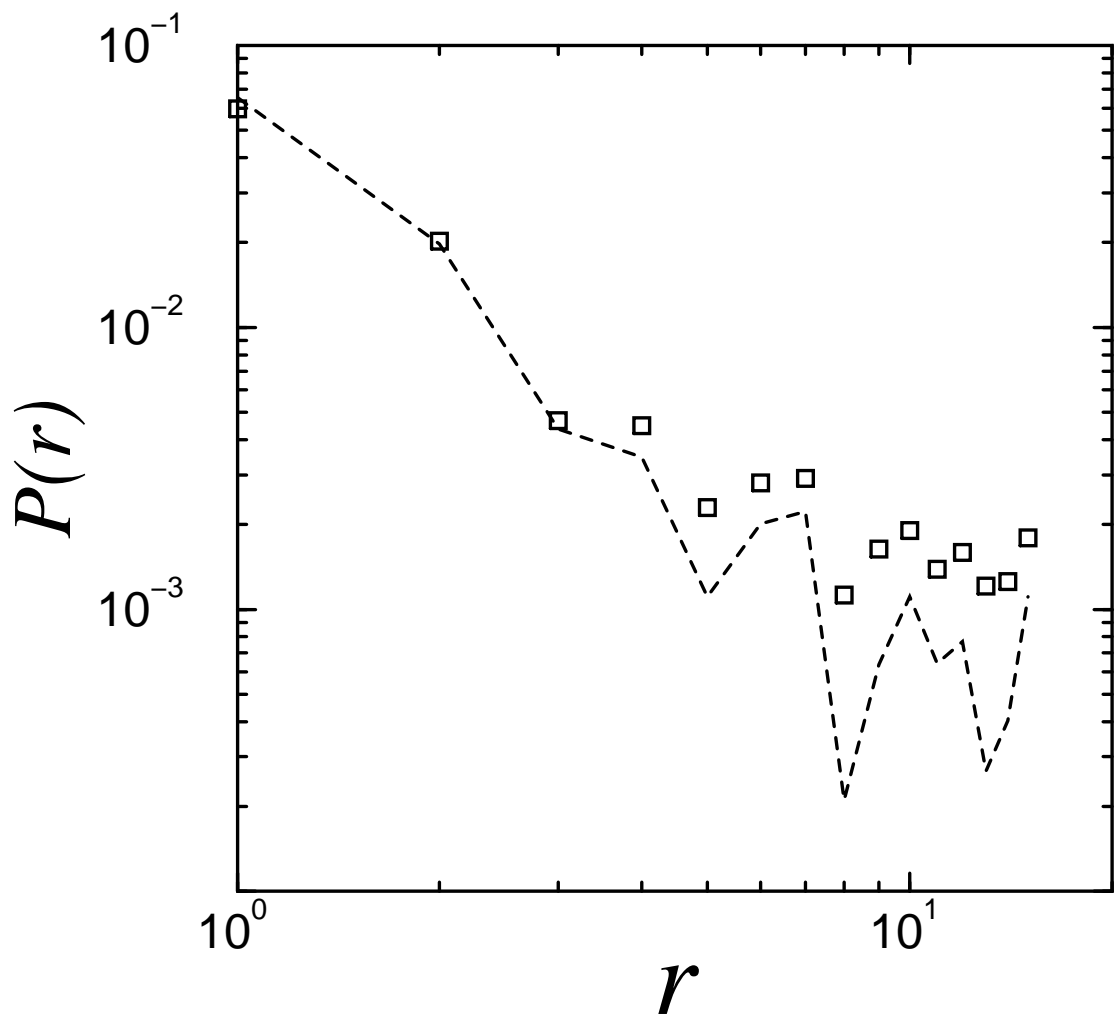
\begin{figure}
\caption{
A similar plot as $\Box$ in Fig.1 except $U=2$ here.
}
\label{fig3}
\end{figure}

In Fig.\ref{fig3}, we display the result for a larger $U=2$.  
We again have
a long-ranged $P(r)$ at large distances, 
although $P(r)$ is slightly reduced from 
the result for $U=1$.
This is consistent with the weak-coupling theory, in which
$K_\rho$ is a {\it decreasing} function of $U$ so that once
the spin gap opens for $U>0$, the paring correlation decays faster
for larger values of $U$. 

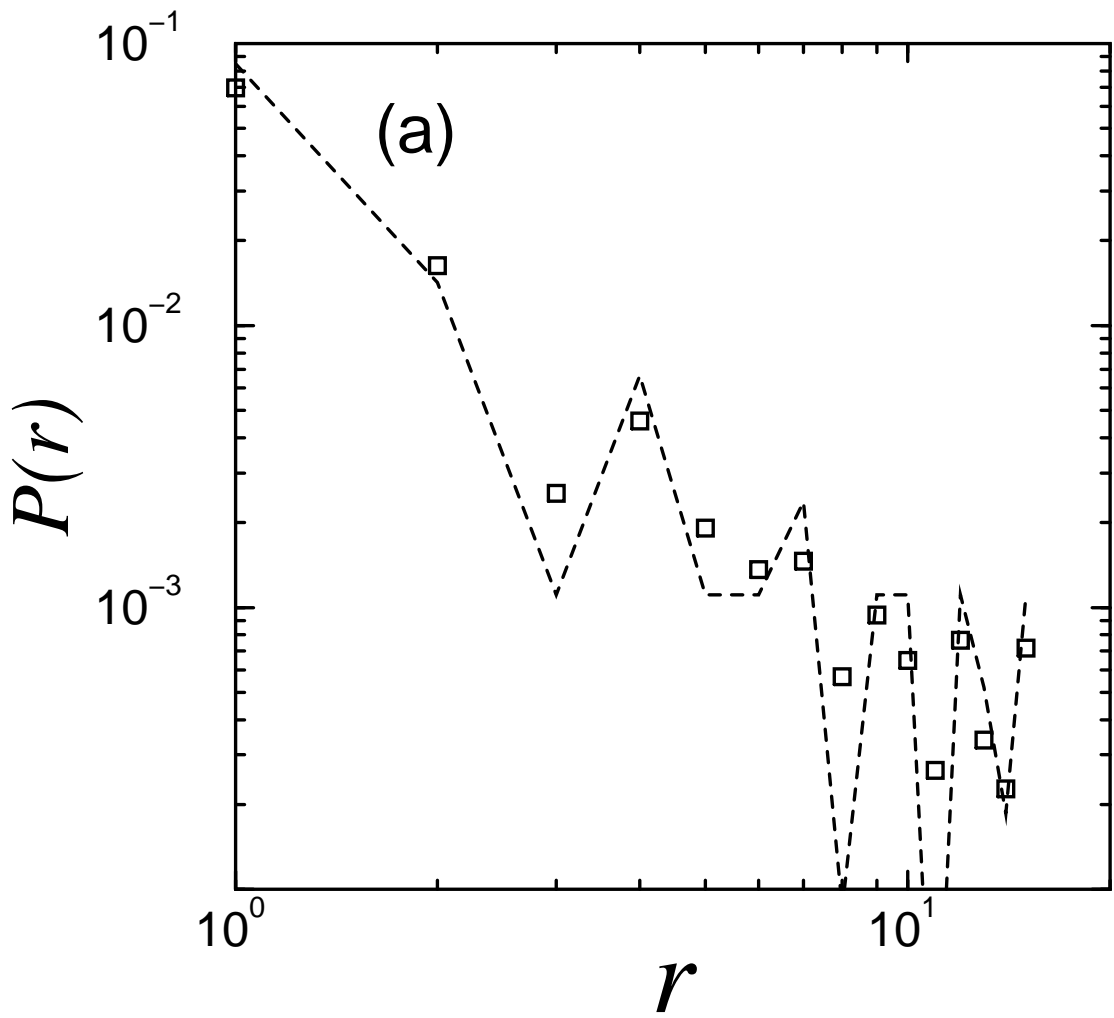
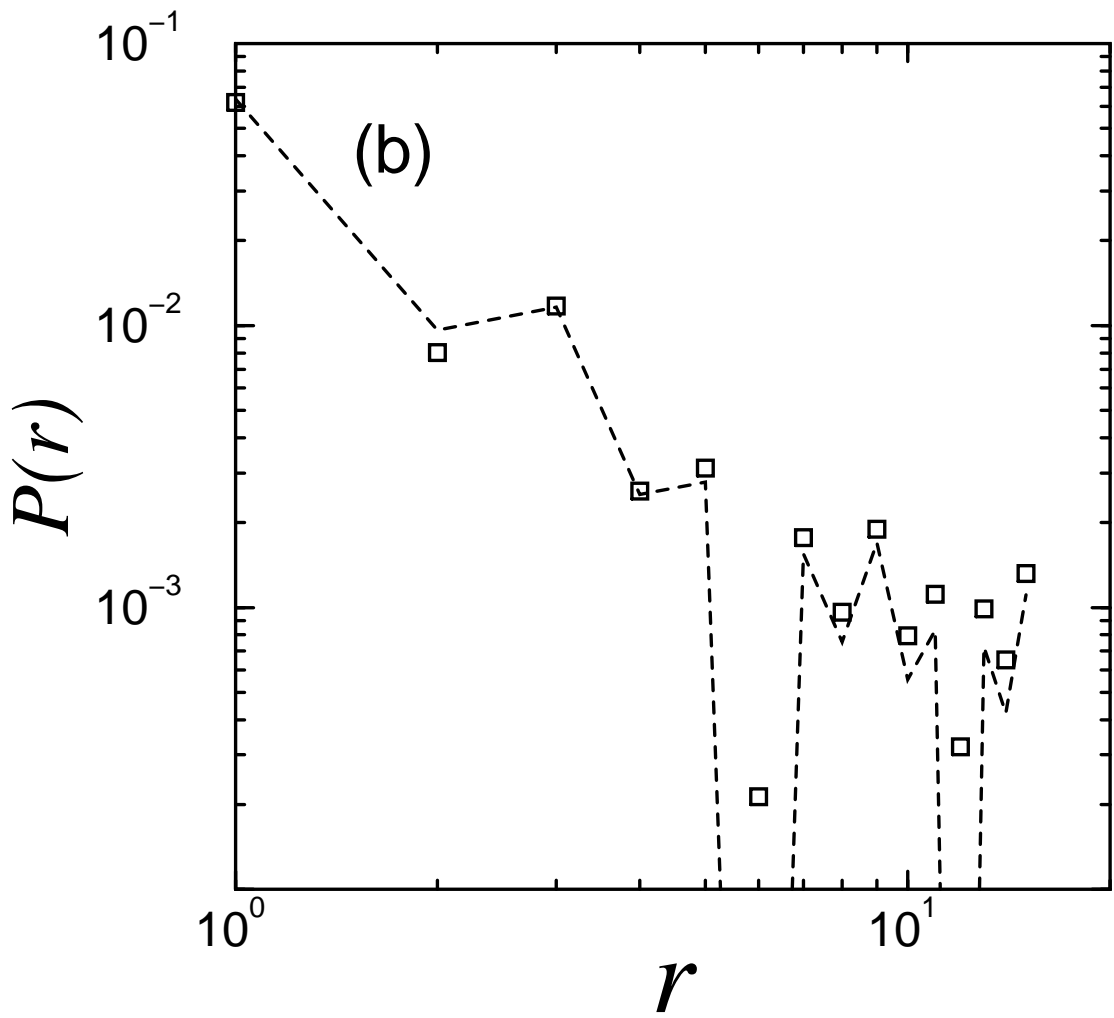
\begin{figure}
\caption{
The pairing correlation $P(r)$ ($\Box$) against $r$ 
for a 30-rung system for $U=2$ with (a) $t_{\perp}=0.99$
and 60 electrons (half-filled), and (b) $t_{\perp}=1.01$ and 40 electrons 
(half-filled bonding band).
The dashed line represents the non-interacting result.
}
\label{fig4}
\end{figure}

To explore the effect of umklapp processes, 
we now turn to the filling dependence for a fixed interaction $U=2$.
We have tuned the value of $t_{\perp}$ to ensure that 
the level offset $(\Delta\varepsilon)$ at the Fermi level 
is as small as O($0.01t$) for $U=0$.
In this way, we can single out the effect of umklapp processes from
those due to large values of $\Delta\varepsilon$.
If we first look at the half-filling (Fig.\ref{fig4}(a)), the decaying form 
is essentially similar to the $U=0$ result.  
At the half-filling interband umklapp processes emerge and, 
according to the weak-coupling 
theory, open a charge gap, which results in an exponential decay of the 
pairing correlation. It is difficult to tell from our data
whether $P(r)$ decays exponentially.  This is probably due to the 
smallness of the charge gap. In fact, Noack {\it et al.}\cite{Noack} 
have obtained with DMRG an exponential decay for larger values of $U$,
for which a larger charge gap is expected.

When $n$ is decreased down to 0.667 (Fig.\ref{fig4}(b)), 
we again observe an absence of enhancement in $P(r)$.  
This is again consistent with the weak-coupling theory\cite{Balents}: 
for this band filling, the number of electrons in the bonding band 
coincides with $N(=30)$ at $U=0$, i.e., the bonding band is half-filled. 
This will then give rise to 
intraband umklapp processes within the bonding band resulting in the 
`C1S2' phase discussed in ref.\onlinecite{Balents}, 
in which the spin gap is destroyed
so that the pairing correlation will no longer decay slowly\cite{DMRG}.

In summary, we have seen that there are three possible 
causes that reduce the pairing correlation function in the 
Hubbard ladder:  
(i) the discreteness of the energy levels, 
(ii) reduction of $K_\rho$ for large values of $U/t$, and 
(iii) effect of intra- and interband 
umklapp processes around specific band fillings.
The first one is a finite-size effect, while the latter two are present
in infinite systems as well. 
We can make a possible interpretation to the existing results 
in terms of these effects.
For 60 electrons on 36 rungs with $t_{\perp}=1.5t$ 
in ref.\onlinecite{Asai}, for instance, 
the non-interacting energy levels have a significant 
offset $\sim 0.15t$ between bonding and anti-bonding levels 
at the Fermi level, which may be the reason 
why the pairing correlation is not enhanced for $U/t=2$.  
For a large $U/t(=8)$ in refs.\onlinecite{Noack,Noack2}, 
(ii) and/or (iii) in the above may possibly be important 
in making the pairing correlation for $t_{\perp}=t$ not enhanced.  
The effect (iii) should be more serious 
for $t_{\perp}=t$ than for $t_{\perp}=1.5t$ because the bonding 
band is closer to the half-filling in the former. On the other hand, 
the discreteness of the energy levels might exert some effect as well, 
since the non-interacting energy levels
for a 32-rung ladder with 56 electrons $(n=0.875)$ 
in an open boundary condition have an offset 
of $0.15t$ at the Fermi level for
$t_{\perp}=t$ while the offset is $0.03t$ for $t_{\perp}=1.5t$.

Finally, let us comment on a possible relevance of the present result 
to the superconductivity reported recently for a cuprate 
ladder\cite{Uehara}, especially for the 
pressure dependence. 
The material is Sr$_{0.4}$Ca$_{13.6}$Cu$_{24}$O$_{41.84}$, 
which contains layers consisting of 
two-leg ladders and those consisting of 1D chains.
Superconductivity is not observed in the ambient pressure, while 
it appears with 
$T_C\sim 10K$ under the pressure of 3 GPa or 4.5 GPa, and 
finally disappears at a higher pressure of 6 GPa. 
This material is doped with holes with the total doping level of 
$\delta=0.25$, where $\delta$ is defined as the deviation of the 
density of electrons from the half-filling.
It has been proposed that at ambient pressure the holes are mostly in the 
chains, while high pressures cause the carrier to transfer 
into the ladders\cite{Kato}. 
If this is the case, and if most of the holes are 
transferred to the ladders at 6 GPa,
the experimental result is consistent with the present picture, since 
there is no enhancement of the 
pairing correlation for $\delta=0$ and $\delta\sim0.3$ due to the umklapp 
processes as we have seen.
Evidently, further investigation especially in the large-$U$ regime
is needed to justify this speculation.

Numerical calculations were done on HITAC S3800/280
at the Computer Center of the University of Tokyo,
and FACOM VPP 500/40 at the Supercomputer Center,
Institute for Solid State Physics, University of Tokyo.
For the former facility we thank Prof. Y. Kanada for a
support in `Project for Vectorized Supercomputing'.
This work was also supported in part by Grant-in-Aid for Scientific
Research from the Ministry of Education, Science, Sports and Culture
of Japan. One of the authors (T.K.) acknowledges the Japan Society for the 
Promotion of Science for a fellowship.

%%%%%%%%%%%%%%%%%%%%  References %%%%%%%%%%%%%%%%%%%%%%%%%%%%%%%%

%%%%%%%%%%%%%%%%%%%% Figure Captions %%%%%%%%%%%%%%%%%%%%%%%%%%%%%%%

%%%%%%%%%%%%%%%%%%%%%%%%%%%%%%%%%%%%%%%%%%%%%%%%%%%%%%%%%%%%%%%%%%%%%

%\end{multicols}
\end{document}